\documentclass[twoside,twocolumn,9pt]{article}
\usepackage{extsizes}
\usepackage[super,sort&compress,comma]{natbib} 
\usepackage[version=3]{mhchem}
\usepackage[left=1.5cm, right=1.5cm, top=1.785cm, bottom=2.0cm]{geometry}
\usepackage{balance}
\usepackage{mathptmx}
\usepackage{sectsty}
\usepackage{graphicx} 
\usepackage{lastpage}
\usepackage[format=plain,justification=justified,singlelinecheck=false,font={stretch=1.125,small,sf},labelfont=bf,labelsep=space]{caption}
\usepackage{float}
\usepackage{fancyhdr}
\usepackage{fnpos}
\usepackage[english]{babel}
\addto{\captionsenglish}{%
  
}
\usepackage{array}
\usepackage{droidsans}
\usepackage{charter}
\usepackage[T1]{fontenc}
\usepackage[usenames,dvipsnames]{xcolor}
\usepackage{setspace}
\usepackage[compact]{titlesec}
\usepackage{hyperref}
\usepackage{ulem}

\usepackage{multirow}

\usepackage{epstopdf}%This line makes .eps figures into .pdf - please comment out if not required.
\definecolor{cream}{RGB}{222,217,201}

\begin{document}

\pagestyle{fancy}
\thispagestyle{plain}
\fancypagestyle{plain}{
%%%HEADER%%%
\renewcommand{\headrulewidth}{0pt}
}
%%%END OF HEADER%%%

%%%PAGE SETUP - Please do not change any commands within this section%%%
\makeFNbottom
\makeatletter
\renewcommand\LARGE{\@setfontsize\LARGE{15pt}{17}}
\renewcommand\Large{\@setfontsize\Large{12pt}{14}}
\renewcommand\large{\@setfontsize\large{10pt}{12}}
\renewcommand\footnotesize{\@setfontsize\footnotesize{7pt}{10}}
\makeatother

\renewcommand{\thefootnote}{\fnsymbol{footnote}}
\renewcommand\footnoterule{\vspace*{1pt}% 
\color{cream}\hrule width 3.5in height 0.4pt \color{black}\vspace*{5pt}} 
\setcounter{secnumdepth}{5}

\makeatletter 
\renewcommand\@biblabel[1]{#1}            
\renewcommand\@makefntext[1]% 
{\noindent\makebox[0pt][r]{\@thefnmark\,}#1}
\makeatother 
\renewcommand{\figurename}{\small{Fig.}~}
\sectionfont{\sffamily\Large}
\subsectionfont{\normalsize}
\subsubsectionfont{\bf}
\setstretch{1.125} %In particular, please do not alter this line.
\setlength{\skip\footins}{0.8cm}
\setlength{\footnotesep}{0.25cm}
\setlength{\jot}{10pt}
\titlespacing*{\section}{0pt}{4pt}{4pt}
\titlespacing*{\subsection}{0pt}{15pt}{1pt}
%%%END OF PAGE SETUP%%%

%%%FOOTER%%%
\fancyfoot{}
\fancyfoot[LO,RE]{\vspace{-7.1pt}\includegraphics[height=9pt]{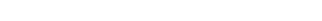}}
\fancyfoot[CO]{\vspace{-7.1pt}\hspace{11.9cm}\includegraphics{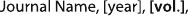}}
\fancyfoot[CE]{\vspace{-7.2pt}\hspace{-13.2cm}\includegraphics{head_foot/RF}}
\fancyfoot[RO]{\footnotesize{\sffamily{1--\pageref{LastPage} ~\textbar  \hspace{2pt}\thepage}}}
\fancyfoot[LE]{\footnotesize{\sffamily{\thepage~\textbar\hspace{4.65cm} 1--\pageref{LastPage}}}}
\fancyhead{}
\renewcommand{\headrulewidth}{0pt} 
\renewcommand{\footrulewidth}{0pt}
\setlength{\arrayrulewidth}{1pt}
\setlength{\columnsep}{6.5mm}
\setlength\bibsep{1pt}
%%%END OF FOOTER%%%

%%%FIGURE SETUP - please do not change any commands within this section%%%
\makeatletter 
\newlength{\figrulesep} 
\setlength{\figrulesep}{0.5\textfloatsep} 

\newcommand{\topfigrule}{\vspace*{-1pt}% 
\noindent{\color{cream}\rule[-\figrulesep]{\columnwidth}{1.5pt}} }

\newcommand{\botfigrule}{\vspace*{-2pt}% 
\noindent{\color{cream}\rule[\figrulesep]{\columnwidth}{1.5pt}} }

\newcommand{\dblfigrule}{\vspace*{-1pt}% 
\noindent{\color{cream}\rule[-\figrulesep]{\textwidth}{1.5pt}} }

\makeatother
%%%END OF FIGURE SETUP%%%

%%%TITLE, AUTHORS AND ABSTRACT%%%
\twocolumn[
  \begin{@twocolumnfalse}
{\includegraphics[height=30pt]{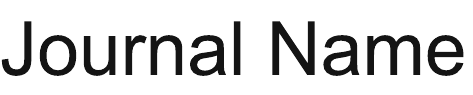}\hfill\raisebox{0pt}[0pt][0pt]{\includegraphics[height=55pt]{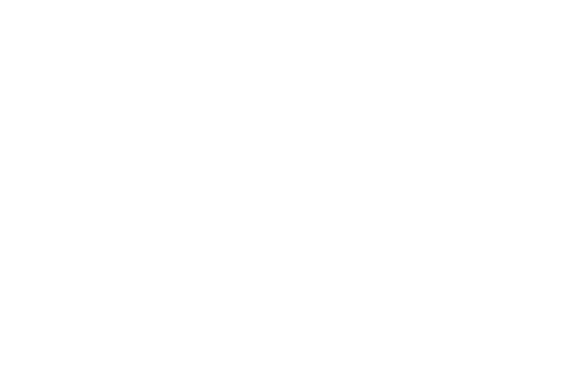}}\\[1ex]
\includegraphics[width=18.5cm]{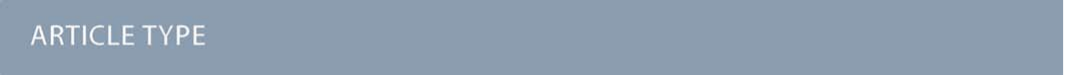}}\par
\vspace{1em}
\sffamily
\begin{tabular}{m{4.5cm} p{13.5cm} }

\includegraphics{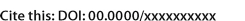} & \noindent\LARGE{\textbf{Silvanite AgAuTe$_4$: a rare case of gold superconducting material$^\dag$}} \\%Article title goes here instead of the text "This is the title"
\vspace{0.3cm} & \vspace{0.3cm} \\

 & \noindent\large{Yehezkel Amiel,$^{\mathsection}$\textit{$^{a}$} Gyanu P. Kafle,$^{\mathsection}$\textit{$^{b}$} Evgenia V. Komleva,\textit{$^{c, d}$} Eran Greenberg,\textit{$^{e, f}$} Yuri S. Ponosov,\textit{$^{c}$} Stella Chariton,\textit{$^{f}$} Barbara Lavina,\textit{$^{f, g}$} Dongzhou Zhang,\textit{$^{f, h}$} Alexander Palevski,\textit{$^{a}$} Alexey V. Ushakov,\textit{$^{c}$} Hitoshi Mori,\textit{$^{b}$} Daniel I. Khomskii,\textit{$^{i}$} Igor I. Mazin,\textit{$^{j}$} Sergey V. Streltsov,\textit{$^{c, d}$} Elena R. Margine,\textit{$^{b}$} and Gregory Kh. Rozenberg$^{\ddag}$\textit{$^{a}$}} \\%Author names go here instead of "Full name", etc.

\includegraphics{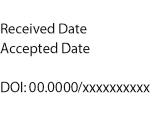} & \noindent\normalsize{
Gold is one of the most inert metals, forming very few compounds, some with rather interesting properties, and only few of them currently known to be superconducting under certain conditions. Compounds of another noble element, Ag, are also relatively rare, and very few of them are superconducting. Finding new superconducting materials containing gold (and silver) is a challenge – especially having in mind that the best high-$T_{\rm c}$ superconductors at normal conditions are based upon their rather close congener, Cu. Here we report combined X-ray diffraction, Raman, and resistivity measurements, as well as first-principles calculations, to explore the effect of hydrostatic pressure on the properties of the sylvanite mineral, AuAgTe$_4$. Our experimental results, supported by density functional theory, reveal a structural phase transition at $\sim$5 GPa from a monoclinic $P2/c$ to $P2/m$ phase, resulting in almost identical coordinations of Au and Ag ions, with rather uniform interatomic distances. Further, resistivity measurements show the onset of superconductivity at $\sim$1.5 GPa in the $P2/c$ phase, followed by a linear increase of $T_c$ up to the phase transition, with a maximum in the $P2/m$ phase, and a gradual decrease afterwards. Our calculations indicate phonon-mediated superconductivity, with the electron-phonon coupling coming predominantly from the low-energy phonon modes. Thus, along with the discovery of a new superconducting compound of gold/silver, our results advance understanding of the mechanism of the superconductivity in Au-containing compounds and dichalcogenides of other transition metals.
} \\

\end{tabular}

 \end{@twocolumnfalse} \vspace{0.6cm}

  ]
%%%END OF TITLE, AUTHORS AND ABSTRACT%%%

%%%FONT SETUP - please do not change any commands within this section
\renewcommand*\rmdefault{bch}\normalfont\upshape
\rmfamily
\section*{}
\vspace{-1cm}

%%%FOOTNOTES%%%
%\footnotetext{\textit{$^{a}$, $^{\ddag}$~School of Physics and Astronomy, Tel Aviv University, 69978 Tel Aviv, Israel. Fax: 03-6409047; Tel: 03-6409047; E-mail: emtsm@tauex.tau.ac.il}}
\footnotetext{\textit{$^{a}$~School of Physics and Astronomy, Tel Aviv University, 69978 Tel Aviv, Israel.}}
\footnotetext{\textit{$^{b}$~Department of Physics, Applied Physics, and Astronomy, Binghamton University-SUNY, Binghamton, New York 13902, USA.}}
\footnotetext{\textit{$^{c}$~M.N. Mikheev Institute of Metal Physics UB RAS, 620137, S. Kovalevskaya str. 18, Ekaterinburg, Russia.}}
\footnotetext{\textit{$^{d}$~Ural Federal University, Mira St. 19, 620002 Ekaterinburg, Russia.}}
\footnotetext{\textit{$^{e}$~Applied Physics Division, Soreq NRC, Yavne 81800, Israel.}}
\footnotetext{\textit{$^{f}$~GSECARS, University of Chicago, Chicago, Illinois 60637, USA.}}
\footnotetext{\textit{$^{g}$~X-Ray Science Division, Advanced Photon Source, Argonne National Lab, 60439, USA.}}
\footnotetext{\textit{$^{h}$~School of Ocean and Earth Science and Technology, University of Hawai’i at Manoa, Honolulu, HI 96822, USA.}}
\footnotetext{\textit{$^{i}$~II. Physikalisches Institut, Universität zu Köln, Zülpicher Straße 77, D-50937 Köln, Germany.}}
\footnotetext{\textit{$^{j}$~Department of Physics and Astronomy, George Mason University, Fairfax, USA.}}
\footnotetext{\textit{$^{\mathsection}$ These authors contributed equally to this work.}}
\footnotetext{\textit{$^{\ddag}$ Corresponding author; Fax: +972-26785301; Tel: +972-504071266; E-mail: emtsm@tauex.tau.ac.il}}

%Please use \dag to cite the ESI in the main text of the article.
%If you article does not have ESI please remove the the \dag symbol from the title and the footnotetext below.
\footnotetext{\dag~Electronic Supplementary Information (ESI) available: technical details about the experimental and calculation methods used; discussion of the Raman spectroscopy and DFT results; additional figures and tables. See DOI: 10.1039/cXCP00000x/}
%additional addresses can be cited as above using the lower-case letters, c, d, e... If all authors are from the same address, no letter is required

%\footnotetext{\ddag~Additional footnotes to the title and authors can be included \textit{e.g.}\ `Present address:' or `These authors contributed equally to this work' as above using the symbols: \ddag, \textsection, and \P. Please place the appropriate symbol next to the author's name and include a \texttt{\textbackslash footnotetext} entry in the the correct place in the list.}

%%%END OF FOOTNOTES%%%

%%%MAIN TEXT%%%%
%The main text of the article\cite{Mena2000} should appear here.

%%%%%%%%%%%%%%%%%%%%%%

\section{Introduction}

Both gold and silver are known as inert metals, which do not easily react with other chemical elements. Among the gold-containing compounds (which are not very common anyway), only a few are superconducting and the situation with the silver-based systems is similar. Famous due to an incommensurate crystal structure, the mineral calaverite, AuTe$_2$~\cite{Dam1985,Streltsov2018}, is a metal, but becomes superconducting~\cite{Kudo2013} under applied pressure or when doped with Pd or Pt.  Some alloys and intermetallic compounds of gold, $e.g.$, Au$_2$Bi~\cite{Haas1931} and  Nb$_3$Au~\cite{Wood1956}  are superconducting, but superconducting compounds combining Au and non-metallic elements are in fact extremely rare.  While some other systems have been theoretically predicted to be superconducting with rather high critical temperatures~\cite{Rahm2017}, we are aware only of one more (besides the above mentioned calaverite AuTe$_2$) compound SrAuSi$_3$, recently synthesised  under high-pressure~\cite{Isobe2014}. Therefore, finding new superconducting compounds of gold (and silver) is a challenge.

Whereas Cu$^{2+}$  is quite stable, Ag and even more so Au are rarely seen in +2 oxidation state, so that typically compounds having nominally Ag$^{2+}$ or Au$^{2+}$ ions tend to disproportionate into Ag$^{1+}$ + Ag$^{3+}$. On the other hand, this very tendency of charge disproportionation could in principle be even more favourable for superconductivity: creating the states with zero and two holes in $d$ shell (like in Ag$^{3+}$ or Au$^{3+}$) reminds of the tendency to form  Cooper pairs, and it is often described theoretically in a ``negative-$U$'' Hubbard-like model, corresponding to an effective electron attraction, which could be beneficial for superconductivity.
 
Keeping all these facts in mind, we undertook the search for novel Au-containing superconductors, and indeed we found superconductivity in a ``close relative" of calaverite AuTe$_2$, in sylvanite, AuAgTe$_4$, which can be considered as calaverite where half of the gold ions are replaced by silver. This rather rare mineral has a somewhat simpler crystal structure than AuTe$_2$: it has a layered structure like in typical dichalcogenides MX$_2$, but, in contrast to AuTe$_2$, which exhibits an incommensurate modulation in the triangular layer of, nominally, Au$^{2+}$ (see the solution of this puzzle in Ref.~\cite{Streltsov2018}), in AuAgTe$_4$ the Au and Ag ions are ordered in a stripy fashion, and they have practically integer valencies  Au$^{3+}$ and Ag$^{1+}$, though, strictly speaking, these notions may be not fully applicable in this case because the material is a metal. Nevertheless, the structural data at ambient pressure correspond to this valence assignment, and crystal chemistry indeed confirms this:
Au and Ag ions sit inside  Te octahedra, which are so strongly distorted that they rather resemble square coordination for Au and dumbbells for Ag, very typical for $d^8$ and $d^{10}$ ions.  Due to the Jahn-Teller effect Au$^{3+}$ with the low-spin $d^8$ configuration and doubly occupied $3z^2-r^2$ orbital almost always prefers strongly elongated octahedra or even square coordination. The linear coordination of $d^{10}$ ions such as Ag$^{1+}$ or Hg$^{2+}$ is attributed to the second-order Jahn-Teller, which describes mixing of completely filled $d$ and empty $s$ states, see, $e.g.$ Ref.~\cite{Orgel1958,Burdett1992}. 

Some of us have theoretically predicted that this structure with strongly distorted Te octahedra is unstable under pressure~\cite{Ushakov2019} and now we can confirm this experimentally: at pressures higher than $\sim$5~GPa the coordination of Ag and Au ions becomes almost identical.  And, even more importantly, at $\sim$1.5~GPa this material becomes superconducting, demonstrating also an abrupt increase of the superconducting critical temperature at the phase transition.

Thus, we have found yet another chemical compound of gold (and silver) that is  superconducting. Our theoretical analysis demonstrated that superconductivity here is likely of a conventional type, predominantly due to electron-phonon interactions.  Eventual contribution of the ``negative-$U$'' mechanism seems to play here a minor role.

\begin{figure}[t]
\includegraphics[width=0.45\textwidth]{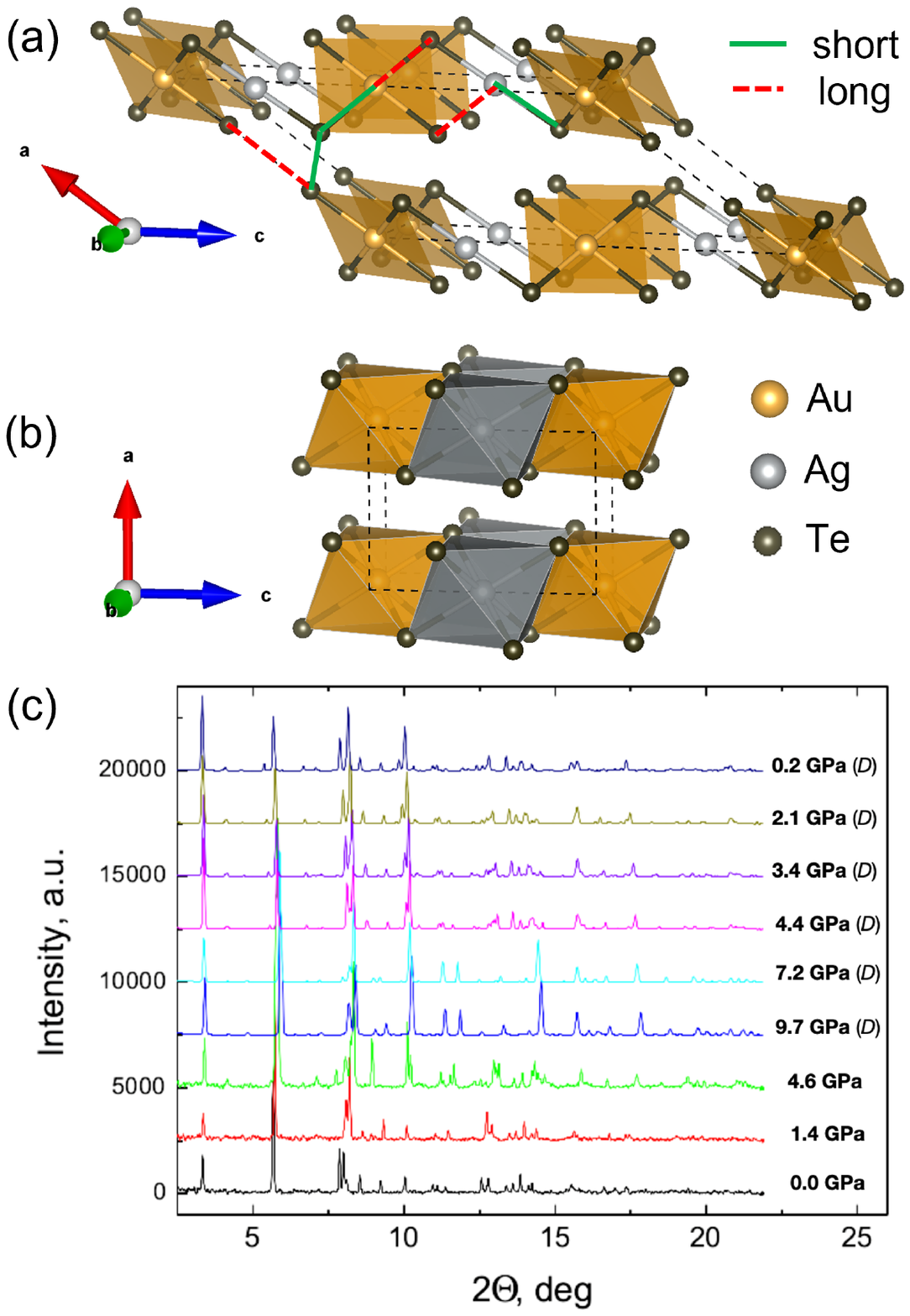}
\caption{\label{F-1a} Polyhedral representation of AuAgTe$_4$ crystal structure of the (a)  low-pressure ($P < 5$~GPa) $P2/c$ phase (the highly distorted octahedra surrounding Au and Ag are shown as square planar and dumb-bell, respectively), and (b) high-pressure ($P > 5$~GPa) $P2/m$ phase drawn using VESTA~\cite{VESTA} software. (c) Pressure evolution of the XRD patterns of AuAgTe$_4$ at compression up to $\sim 9.7$~GPa and following decompression ($D$) to 0.2~GPa ($\lambda_{x-ray}$ = 0.2952 \AA).}
\end{figure}

%%%%%%%%%%%%%%%%%%%%%%%%%

\section{Experimental and theoretical details}

\subsection{Samples and characterization}
The experiments were performed with high-quality natural single-crystals of silvanite, AuAgTe$_4$\footnote{The XRD and resistance $R(P,T)$ experiments were performed with a natural single-crystal of silvanite from the classical Transylvania locality from the private collection of Ladislav Bohat\'y and Petra Becker-Bohat\'y, University of Cologne. The Raman experiment was carried out using natural single-crystals of silvanite from the Kochbulak deposit, Kuraminsky Range, Uzbekistan. XRD results for both samples are in agreement with each other.}.  Custom diamond anvil cells (DACs) and DACs of symmetric design were used to induce high pressure, with Ne or KCl serving as a pressure-transmitting medium. The pressure was determined using the ruby R1 fluorescence line as a pressure marker, as well as the Ne unit-cell volume in the case of X-ray diffraction (XRD) studies. Single-crystal (SC) XRD experiments were performed at the 13-ID-D beamline (mainly) and the beamline 13-BM-C of the APS synchrotron (Argonne, IL, USA). Electrical resistance measurements were performed as a function of pressure and temperature using the standard four-probe method. 

\subsection{Density functional theory calculations}

The \textsc{Quantum Espresso}~\cite{QE} package was used to perform first-principles calculations within the density functional theory (DFT), while the superconducting properties were investigated using the \textsc{EPW} code~\cite{Giustino2007, EPW, Margine2013}.

For further information on the experimental and theoretical methods see the Supplementary Information$^\dag$. 

%%%%%%%%%%%%%%%%%%%%%%%%%%%

\section{\label{sec:expt-results} Experimental results}

\subsection{X-ray diffraction}

XRD patterns obtained up to about 10~GPa are shown in Fig.~\ref{F-1a}(c) (see also Table~S1$^{\dag}$). Low pressure (LP) patterns up to $\sim$5~GPa can be identified as a monoclinic $P2/c$ structure~\cite{Tunnell1952}, where both Ag and Au are octahedrally coordinated by Te, but these octahedra are so strongly distorted that in fact Ag has a dumb-bell and Au  a square planar surrounding, as shown in Fig.~\ref{F-1a}(a). Each Te atom is surrounded by three Au or Ag atoms and three Te atoms, and it is much closer to one of its three Te neighbors than to the other two. Therefore, there are two types of Te-Te interlayer bonds labeled as short and long in Fig.~\ref{F-1a}(a). All the Te atoms are thus members of well-defined Te$_2$ clusters.

\begin{figure}[t]
\includegraphics[width=0.5\textwidth]{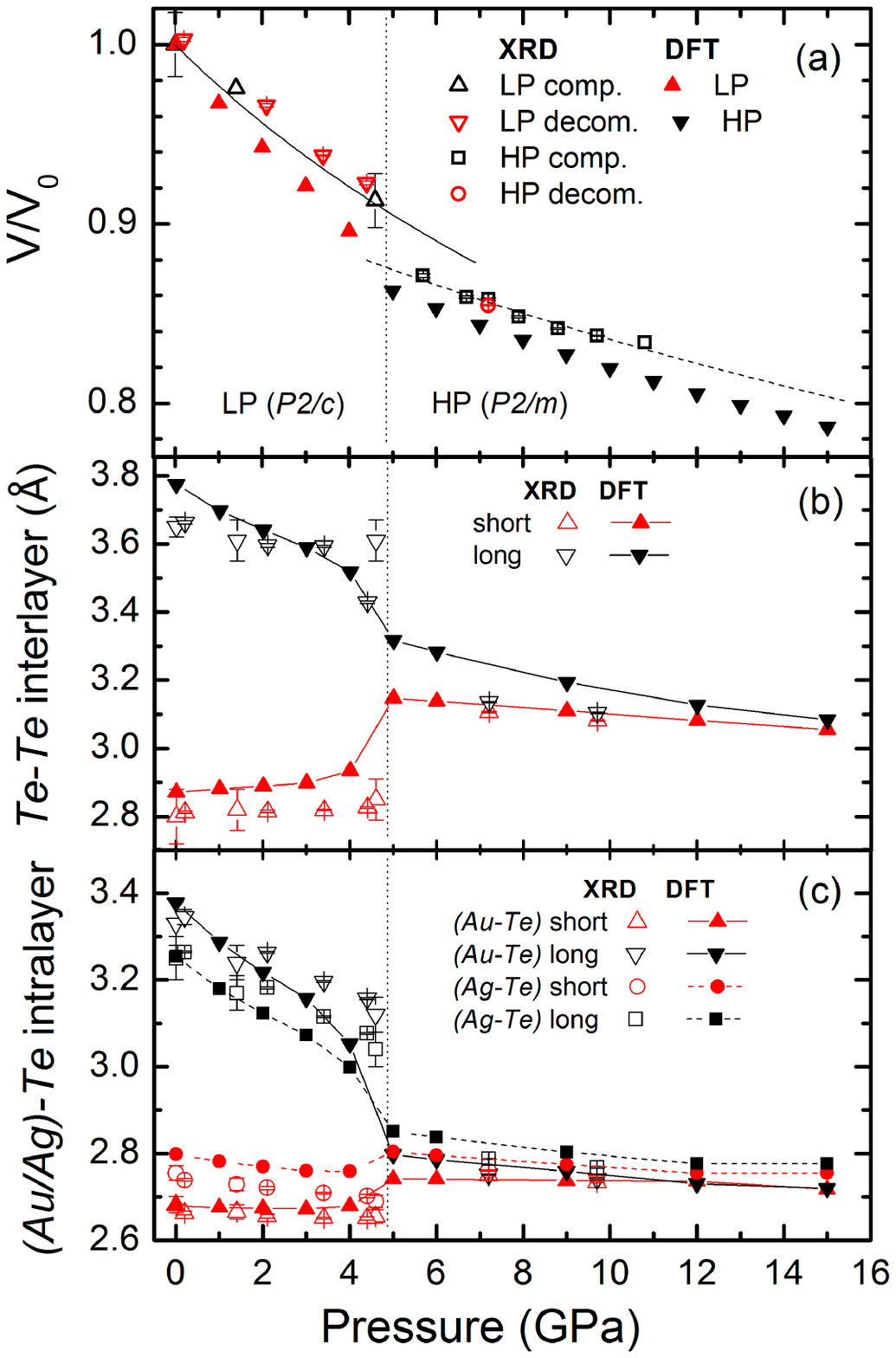}
\caption{\label{F-2a} (a) Pressure dependence of the unit-cell volume of AuAgTe$_4$ determined in the single crystal and powder XRD experiments (open symbols), while fits by Birch-Murnaghan equation of state~\cite{Anderson1995} are shown by lines. DFT results are presented by filled symbols. Experimentally observed (open symbols) and calculated in DFT (lines and filled symbols) pressure dependence of the (b) Te-Te interlayer and (c) (Au/Ag-Te) intralayer distances of AuAgTe$_4$.}
\end{figure}

At $\sim$5~GPa, an onset of a new high-pressure (HP) phase is observed (Fig.~\ref{F-1a}(c), Fig.~S1$^{\dag}$), whose XRD patterns could be fitted well with the more layered $P2/m$ structure (see Tables S2, S3$^{\dag}$). In the HP phase the Te$_6$ octahedra around Au and Ag become regular and practically identical, and the Te-Te interlayer distances become almost equal (Fig.~\ref{F-1a}(b), Fig.~\ref{F-2a}(b,c)). We note that this phase transition is in excellent agreement with the recent theoretical prediction and present DFT calculations~\cite{Ushakov2019}. 

The LP phase $V(P)$ data can be fit well with a second-order Birch-Murnaghan equation of state (BM2 EOS)~\cite{Anderson1995} as shown in Fig.~\ref{F-2a}(a), resulting in $V_0=335.3(11)$~\AA$^3$ and $K_0=41.1(24)$~GPa, where $K_0$ and $V_0$ are the bulk modulus and the unit-cell volume at 1 bar and 300 K, respectively, with the bulk modulus first derivative fixed at $K' = 4$. For the HP phase, the performed fit using the BM2 EOS results in $V_0=307.8(25)$~\AA$^3$ and $K_0=81(9)$~GPa (combining both the SC refinements, and the wide images, which were collected during continuous rotation within a single exposure and were analyzed as if they were ``powder'' data for the HP phase). Close to the transition pressure, at 5 GPa, the unit-cell volume and bulk modulus are $V=303.5(8)$~\AA$^3$ and $K=60(3)$~GPa, and $V=291.3(8)$~\AA$^3$ and $K=101(9)$~GPa for the LP and HP phases, respectively. Thus, the phase transition is accompanied by a lattice volume contraction of $\sim$ 4\% and a significant increase of the bulk modulus.

\subsection{Raman spectroscopy}

At room temperature and ambient pressure we observe almost all allowed Raman active vibrations for AuAgTe$_4$ ($P2/c$ space group): 7 A$_g$ and 7 weaker B$_g$ phonon modes in the 40--160 cm$^{-1}$ range. Tab. I shows that the experimental and DFT calculated Raman frequencies are in overall good agreement. 

\begin{figure}[t]
\includegraphics[width=0.45\textwidth]{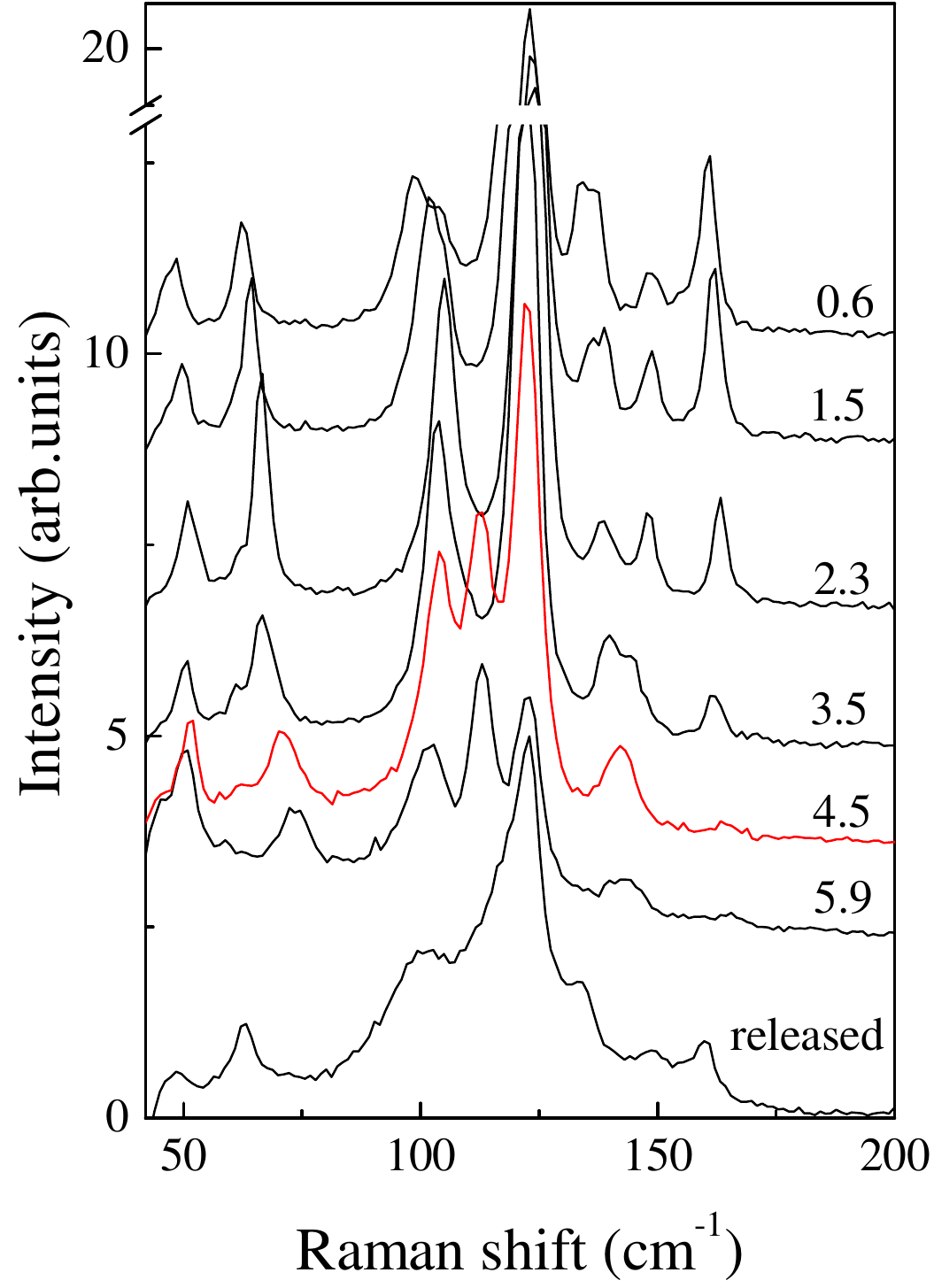}
\caption{\label{press} Raman spectra of AuAgTe$_4$ obtained at different pressures with 633 nm excitation in the polarized geometry.  Numbers in the figure refer to pressure in GPa.}
\end{figure}

\begin{table}[h]
\caption{\label{table1} Comparison of the experimental and calculated Raman active modes of AuAgTe$_4$ at ambient pressure (in cm$^{-1}$). Note that only 7 out of 8 $B_g$ modes have been resolved experimentally. Their frequency values can be determined with an accuracy of about $\pm1$ cm$^{-1}$ using Lorentz fits.}
\begin{tabular}{c c c c c c c c c c}
\hline
\hline
%\multirow{4}{*}
%{\begin{tabular}
\multirow{2}{*}{A$_g$} & Expt. & 47 & 61 & 95 & 102 & 121 & 132 & 158 &     \\ \cline{2-10} 
                       & Calc. & 48 & 60 & 87 & 98  & 112 & 124 & 147 &     \\ \cline{1-10} 
\multirow{2}{*}{B$_g$} & Expt. & 50 & 58 & -- & 84  & 88  & 114 & 134 & 147 \\ \cline{2-10} 
                       & Calc. & 46 & 52 & 59 & 83  & 112 & 129 & 135 & 147 \\ 
\hline
\end{tabular}
\end{table}

 One can see that the Raman spectroscopy clearly detects structural changes above 4~GPa, see Fig.~\ref{press} and Fig.~S4$^{\dag}$. In particular, new modes are observed at 113 and 142 cm$^{-1}$ and the intensity of some lines (158  cm$^{-1}$) changes considerably in the polarized spectra. The polarized spectra in Fig.~S3$^{\dag}$ additionally  support a structural transition in the 4-6~GPa region. In addition to the appearance of new lines in the spectrum, the frequencies of a number of lines either increase significantly (61, 133 and 158 cm$^{-1}$) or decrease (147 cm$^{-1}$) with increasing pressure, the energies of others change nonmonotonically (102 cm$^{-1}$) or increase insignificantly (47 and 121 cm$^{-1}$) (Fig.~S4$^{\dag}$).

It is interesting that there are 7 Raman lines above the  transition. This number is larger than what follows from the selection rules for the refined HP structure ($P2/m$ space group), where should be only 6 Raman-active modes: 4A$_g$ + 2B$_g$. It is well known that Raman spectra provide information not only on the long-range order (since the number of observed lines is determined by the space group of the crystal), but also on the short-range order, being sensitive to local structural distortions. Thus, the appearance of extra lines in the Raman spectra, perhaps, evidences formation of two phases in this pressure range, which is in line with the resistivity measurements discussed below. In addition, one may expect defects in natural crystal, which ensure leakage of a symmetry-forbidden line in the spectrum. In all three experiments, we obtained a somewhat broadened spectrum, compared to the initial crystal, after pressure release (Fig.~\ref{press}).
% We note that with further increase in pressure a significant broadening of most of the lines occurs, which also indicates possible structural disorder. In all three experiments, we obtained a somewhat broadened spectrum, compared to the initial crystal, after pressure release (which was confirmed by polarization measurements shown in Fig.~S3$^{\dag}$). 
%\textcolor{green}{We note that with further pressure increase a significant broadening of most of the lines occurs and we also obtained a somewhat broadened spectrum, compared to the initial crystal, after pressure release.}

\subsection{Resistance measurements}

\begin{figure}[t]
\includegraphics[width=0.45\textwidth]{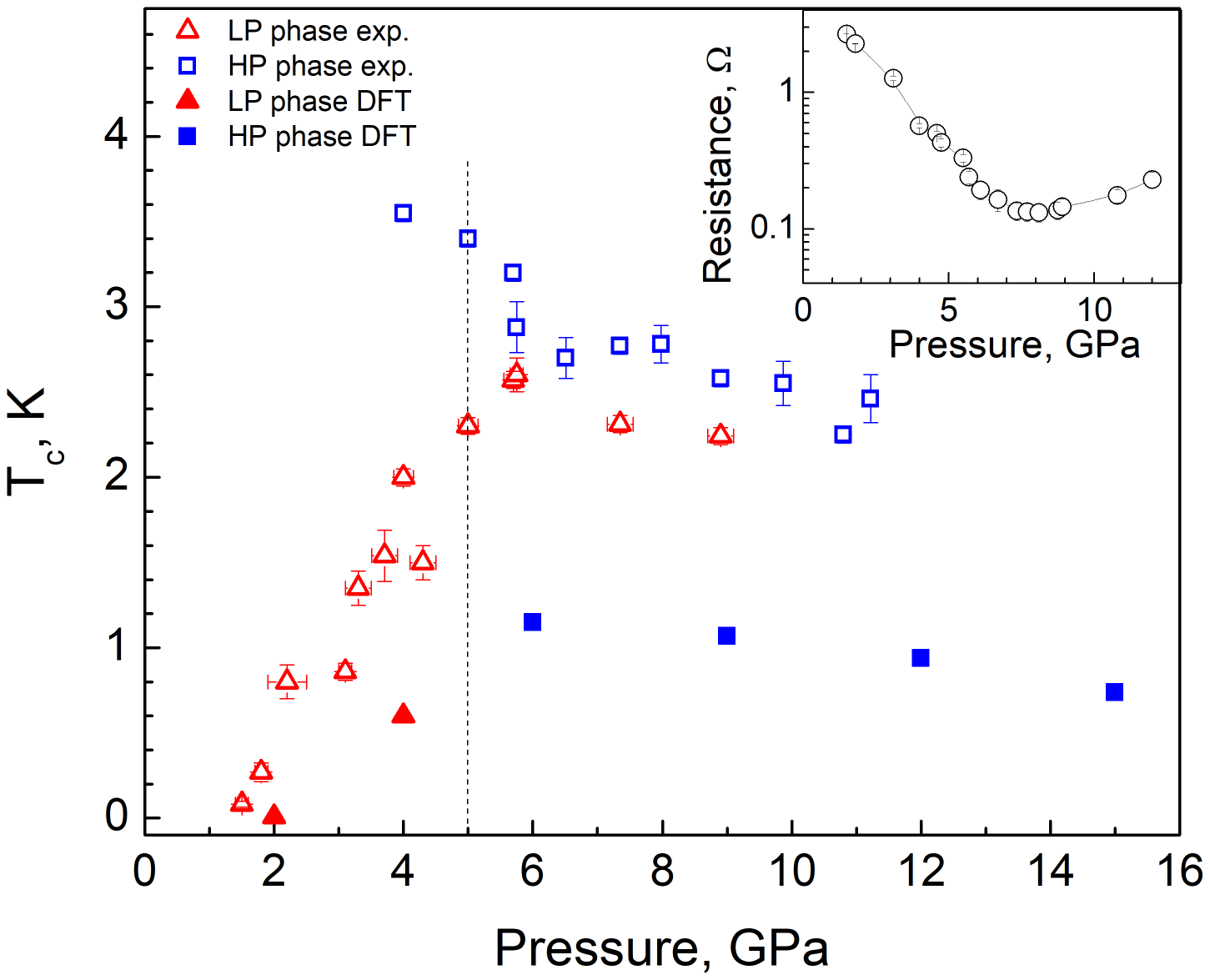}
\caption{\label{F-5} Critical temperature as a function of pressure. Experimental (calculated) results are presented by open (filled) symbols. The vertical line represents the phase separation between the LP and HP phase. The pressure dependence of the room temperature resistance is shown in the inset.}
\end{figure}

The resistivity value of AuAgTe$_4$ at ambient conditions was estimated to be $\sim 3\times10^{-6}$ $\Omega$m, typical of a bad metal. Under pressure AuAgTe$_4$ shows a significant, about a factor of 30, drop in the resistance when it is compressed up to $\sim 7$~GPa, followed by slight increase above this (see the inset of Fig.~\ref{F-5}).

\begin{figure}
\includegraphics[width=0.5\textwidth]{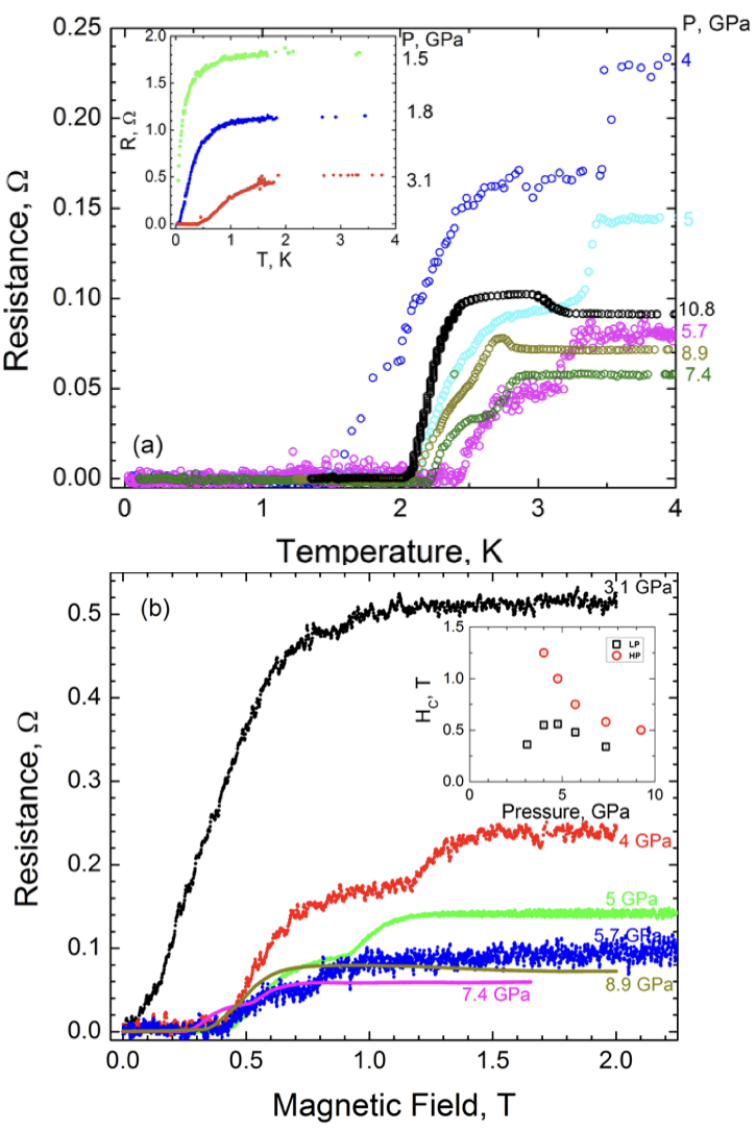}
\caption{\label{F-3a} (a) Temperature dependence of the resistance in AuAgTe$_4$. The graph shows complete superconductivity transitions (resistance drops to zero). The double transitions observed in the $4 < P < 9$~GPa pressure range  are interpreted as a mixture of LP and HP phases. (b) Pressure dependence of the critical field of AuAgTe$_4$. Resistance as a function of magnetic field at different pressures: from 4 to 7.4~GPa two transitions are observed. The inset shows variation of the critical field as a function of pressure.}
\end{figure}

In Fig.~\ref{F-3a}(a)  we show the resistance vs. temperature dependence at various pressures for the most representative run 3 of measurements. One can see that the appreciable drop in resistance coincides with the onset of superconductivity at $P\approx$1.5~GPa with a superconducting critical temperature of $T_c \approx 80$ mK. With further pressure increase, $T_{\rm c}$ increases almost linearly up to $\approx 2.6$ K at 5.7~GPa and then decreases slowly, demonstrating a non-monotonous dome-like shape (Fig.~\ref{F-3a}(a)). Thus, we can conclude that the onset of superconductivity takes place in the LP phase of AuAgTe$_4$ at pressures above $\sim$1.5 GPa. With this, starting from $P = 4$~GPa one can clearly see a drastic change in $R(T)$ behavior: the $R(T)$ curves have two distinct transitions, signifying the appearance of an additional phase with a higher transition temperature ($T_c \approx 3.5$ K at $P = 4$~GPa). As demonstrated by our XRD data, in this pressure range a crystallographic phase transition occurs forming the HP $P2/m$ phase. We can, therefore, interpret the double transition as coexistence of the LP and the HP phases, both being superconducting with a higher $T_{\rm c}$ for the HP phase. The critical temperature for the HP phase decreases appreciably with pressure, approaching the $T_{\rm c}$ of the LP phase. Above $\sim 9$~GPa a single transition is observed. 

In Fig.~\ref{F-3a}(b), we show magnetoresistance measurements at 2 K for run 3. From these curves we can extract the upper critical magnetic field $H_c$ as the field at which the resistance is half of the normal state resistance. In the pressure range from 4 to 7.4~GPa, where the temperature dependence exhibits two transitions, two transitions are observed in magnetic field. These measurements are consistent with our interpretation that both structural phases coexist within this pressure range. The corresponding upper critical field is plotted in the inset of Fig.~\ref{F-3a}(b). Our definition of $H_c$ is not appropriate for the pressure range where we observe a double transition (coexistence regime). In the latter case we estimated $H_c$ as the mid-point value of each transition for each phase. The critical magnetic field was found to depend on pressure, as it varies between $\sim 0.2$ and 0.5~T for LP phase, and between $\sim 1.2$ and 0.5 T for the HP phase. We note that, similar to $T_{\rm c}$, the critical field demonstrates a non-monotonous dome-like shape in the LP phase and an appreciable decrease with pressure in the HP phase.

%%%%%%%%%%%%%%%%%%%%%%%%%%

\section{Computational results}

\begin{figure*}[t]
\includegraphics[width=1\textwidth]{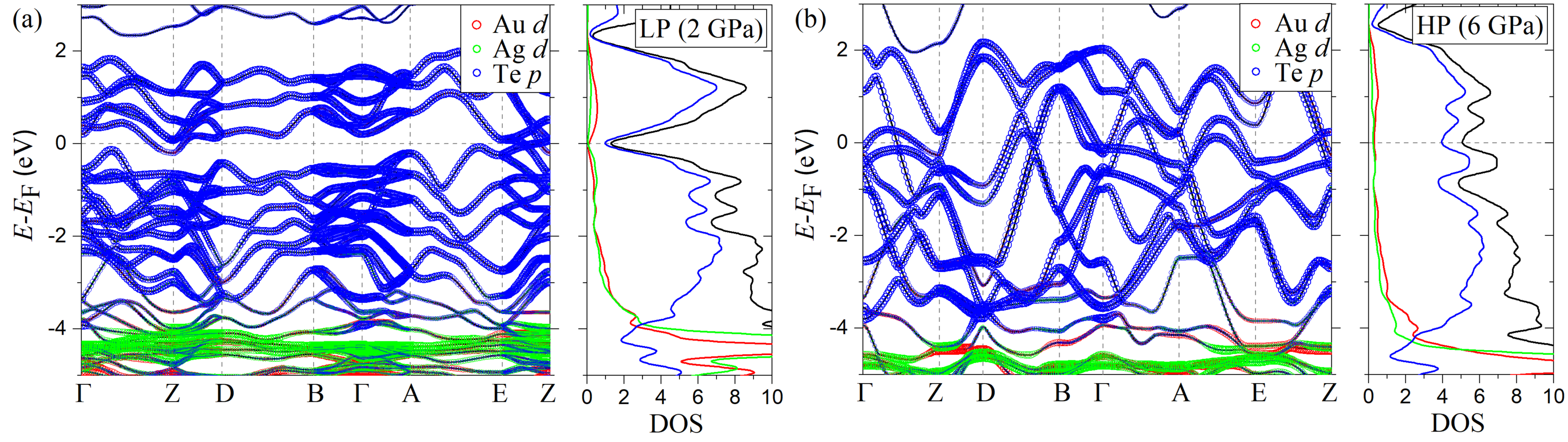}
\caption{\label{F-9} Calculated band structure and density of states (DOS; in states/eV/f.u.) of AuAgTe$_4$ at (a) 2~GPa (low-pressure phase) and  (b) 6~GPa (high-pressure phase). The size of the markers is proportional to the contribution of each orbital character. The solid black line in the DOS panel represents the total DOS and the red, green, and blue lines are the contributions to the DOS from the Au, Ag, and Te atoms, respectively.}
\end{figure*}

As discussed in previous sections, the DFT calculations describe very well the structural transition under pressure (and in fact had predicted this transition \cite{Ushakov2019}). 
The calculated pressure dependence of the crystal volume, lattice parameters and interatomic distances are in good agreement with the experimental results (see Fig.~\ref{F-2a} and Fig.~S2$^{\dag}$).

\begin{figure}[t]
\includegraphics[width=0.5\textwidth]{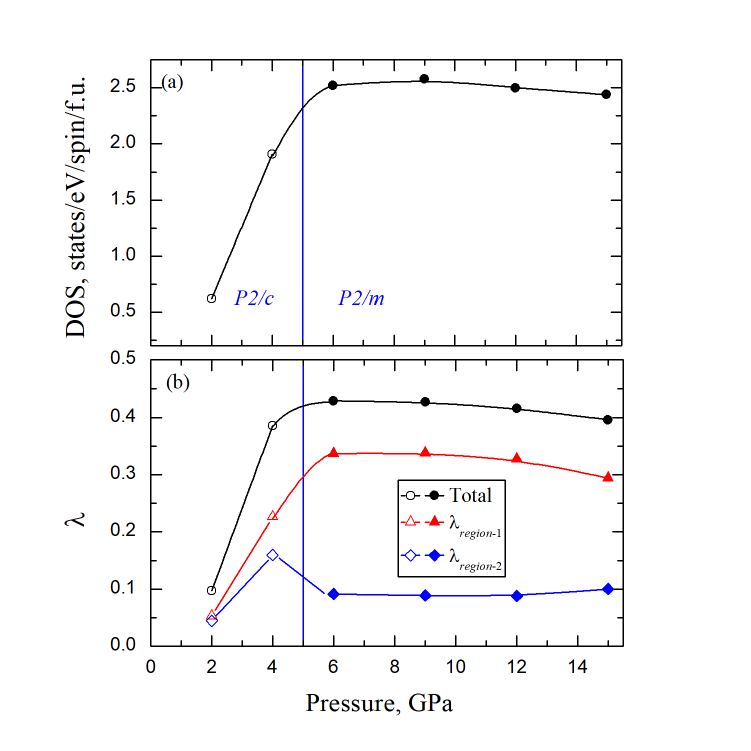}
\caption{\label{F-11} Calculated (a) DOS (states/eV/spin/f.u.) at the $E_{\rm F}$ and (b) total e-ph coupling $\lambda$ and partitioning of $\lambda$ for two frequency regions of AuAgTe$_4$ as a function of pressure. The region-1 and region-2 are the separations in the phonon spectrum below and above 11 meV in the LP phase and 15 meV in the HP phase. The vertical blue line represents the phase separation between the LP and HP phases. The open and closed symbols are for the LP and HP phases, respectively.}
\end{figure}

\subsection{Electronic properties}

The band structure and density of states (DOS)  of AuAgTe$_4$ for the LP phase at 2~GPa and the HP phase at 6~GPa are shown in Fig.~\ref{F-9}, and those at other pressures are given in the Supplemental Fig.~S5$^{\dag}$. It was shown in \cite{Ushakov2019} that the Te-Te dimerization at ambient pressure opens a pseudogap due to the bonding-antibonding splitting, but this pseudogap gradually closes under pressure.
The largest contribution to the DOS at the Fermi level ($E_{\rm F}$) is provided by the Te $p$ states. As pressure increases to 4~GPa (Supplemental Fig.~S5(a)$^\dag$), an electron-like band related to the Te $p$ orbitals lowers in energy along the $\Gamma$-$Z$-$D$ direction, while two hole-like bands along the $\Gamma$-$Z$ and $\Gamma$-$A$-$E$ directions raise in energy and cross $E_{\rm F}$. These changes cause a drastic increase in the total DOS at the Fermi level ($N_{\rm F}$), as shown in Fig.~\ref{F-11}(a) (also see Supplemental Fig.~S5(a)$^\dag$). After the phase transition, the dispersion of all bands is greatly increased, while the major contribution to the DOS remains to be from the states of Te $p$ character (Fig.~\ref{F-9}(b) and Fig.~S5(b-d)$^\dag$). At 6~GPa, the $N_{\rm F}$ of the HP phase is 31\% of that in the LP phase at 4~GPa. No significant changes are observed in the band structure of the HP phase in the considered 6-15~GPa range, leaving $N_{\rm F}$  nearly constant (see Fig.~\ref{F-11}(a)). Overall, we observe that the Te $p$ states provide almost 75\% of the total DOS at $E_{\rm F}$ for all pressure points. We also checked the effect of spin-orbit coupling and found that the change in the total DOS near the Fermi level is negligible for both the low- and high-pressure phase (see Supplemental Fig.~S7(a)$^\dag$).

\subsection{Phonons}
\begin{figure*}[tb]
\includegraphics[width=\textwidth]{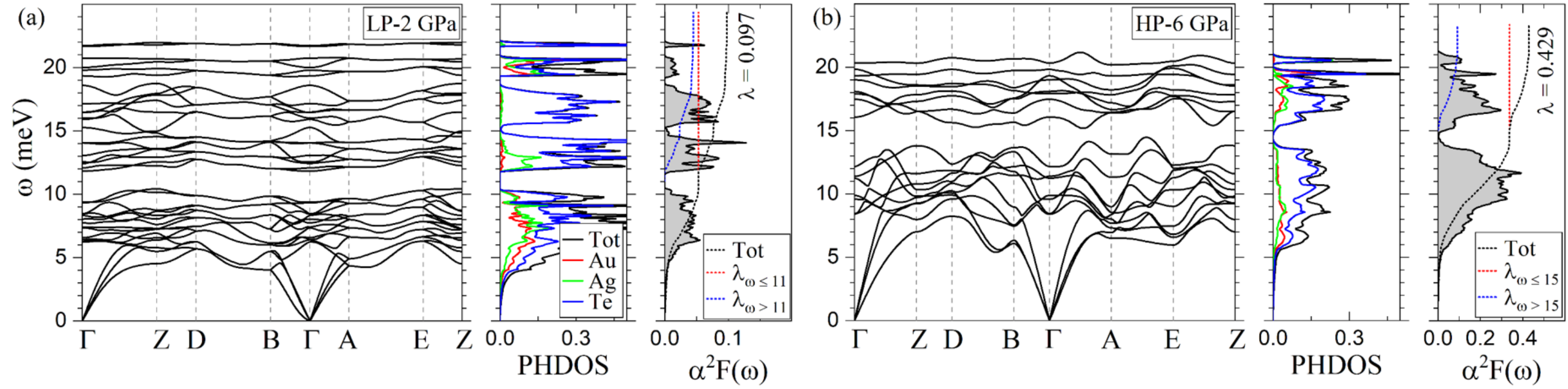}
\caption{\label{F-10} Calculated phonon dispersion, phonon density of states (PHDOS), and Eliashberg spectral function $\alpha^2F(\omega)$ of AuAgTe$_4$ at (a) 2 GPa (low-pressure phase) and  (b) 6~GPa (high-pressure phase).}
\end{figure*}

Figure~\ref{F-10} shows the phonon dispersion, the phonon density of states (PHDOS), the isotropic Eliashberg spectral function $\alpha^2F(\omega)$, and the cumulative electron-phonon (e-ph) coupling ($\lambda$) of AuAgTe$_4$ for the LP phase at 2~GPa and the HP phase at 6~GPa, and the other pressure points are shown in Supplemental Fig.~S6~$^\dag$. We find that the LP phase is stable at 2 and 4~GPa, but dynamically unstable at 6 GPa. As displayed in Fig.~\ref{F-10}(a), the phonon spectrum of the LP phase at 2~GPa is divided into two regions (region-1 and region-2) separated by a large gap  around 11~meV. The PHDOS decomposed according to atomic contributions shows that the Te-derived modes extend over the whole spectrum. On the other hand,  the Au and Ag contributions to the PHDOS are dominant in region-1, while there is no contribution from either Au or Ag in the 14-18~meV range. In addition, the optical branches above 11~meV are not as dispersed as those below this threshold.  At 4~GPa (Supplemental Fig.~S6(a)~$^\dag$), the frequency gap around 11~meV  is closed as one optical branch couples to the modes in the lower region along the $B$-$\Gamma$-$A$ direction. Another noticeable aspect is the softening of the low-energy acoustic branches along the $D$-$B$ direction.

The HP phase is found to be dynamically stable in the 6-15~GPa pressure range considered in our study. The major contribution to the PHDOS comes from the Te vibrations, but the spectrum lacks the phonon branches solely related to the Te vibrations present in the LP phase in the 14-18~meV range, since now the long and short Te-Te interlayer distances are very close to each other (Fig.~\ref{F-2a}(b)). In addition, the vibrational modes associated with the Au and Ag atoms harden under pressure as shown in the PHDOS (Fig.~\ref{F-10}(b) and Supplemental Fig.~S6(b-d)$^\dag$). The low- and high-frequency regions (region-1 and region-2) remain separated by a small gap centered around 15-16~meV at all pressure points. As in the LP phase, the optical phonon branches are hardening under compression, while the low-energy acoustic modes along the $D$-$B$ direction and at the $E$-point soften.

\subsection{Superconductivity}
\begin{figure*}[tb]
	\centering
	\includegraphics[width=0.95\linewidth]{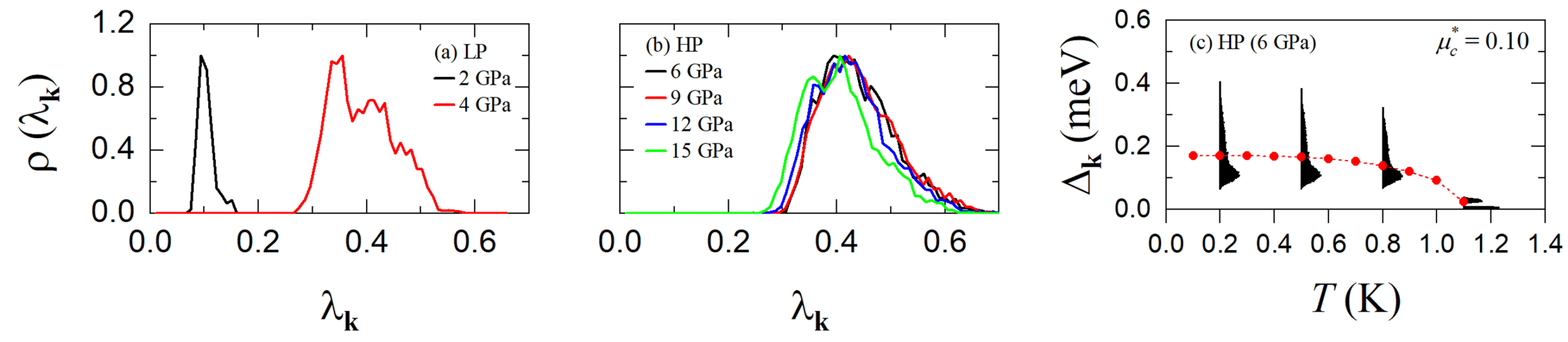}
	\caption{\label{F-9} Distribution of the e-ph coupling strength $\lambda_{\textbf{k}}$ of (a) the LP and (b) the HP phase of AuAgTe$_4$ at various pressures. (c)  Energy distribution of the anisotropic superconducting gap $\Delta_\textbf{k}$ of the HP phase as a function of temperature at 6~GPa. The red symbols represent the isotropic superconducting gap at 6 GPa.  The Coulomb parameter is set to $\mu_c^*=0.10$.}
\end{figure*}

In order to investigate superconducting properties, we first evaluated the isotropic Eliashberg spectral function $\alpha^2F(\omega)$ and the cumulative e-ph coupling strength $\lambda(\omega)$. As depicted in Fig.~\ref{F-10}(a), for the LP phase at 2~GPa, the low-frequency modes below 11~meV associated with vibrations from all atoms contribute 55\% to the total e-ph coupling $\lambda$ = 0.097. An almost four fold increase in the total e-ph coupling ($\lambda$ = 0.386) is found at 4~GPa as the DOS at the Fermi level rises by a similar factor as shown in Fig.~\ref{F-11}. From the division of the phonon spectrum, we found that region-1 below 11~meV supplies 59\% of $\lambda$, slightly more than at 2~GPa since now the coupling due to the acoustic modes has strengthened with the phonon softening.  

In the HP phase at 6~GPa, the low-energy phonons below 15~meV (region-1) make up approximately 80\% of the total $\lambda$ = 0.429  (Fig.~\ref{F-10}(b)).  A comparative analysis of the $\alpha^2F(\omega)$ in the two phases shows that the HP phase lacks in region-2 (above 15~meV) the coupling coming solely from the Te-derived vibrations in the LP phase (14-18~meV range), leading to a factor of two reduction in $\lambda$ in the upper frequency region. Under further compression, the ratio of the low- to high-frequency phonons contribution to the total e-ph remains nearly constant and $\lambda$  decreases slowly following the same trend as the total DOS at the Fermi level. 

Finally, to estimate the superconducting critical temperature ($T_{\rm c}$), we solved the isotropic Migdal-Eliashberg equations implemented in the \textsc{EPW} code~\cite{Giustino2007, EPW, Margine2013} using a Coulomb pseudopotential $\mu_c^* = 0.10$. Fig.~\ref{F-5} shows the calculated $T_{\rm c}$ and its comparison with the experiment. For the LP phase, we estimate $T_{\rm c}$ to be very close to $zero$ at 2~GPa, consistent with the experiments at 1.8~GPa, with an increase to 0.6~K at 4~GPa. After the phase transition, we obtain a maximum $T_{\rm c}$ of 1.2~K at 6~GPa for the HP phase. In line with the resistivity measurements, the superconducting critical temperature decreases slowly in response to pressure, as shown in Fig.~\ref{F-5}. Altogether, we find that the dome-shape behavior of the $T_{\rm c}$ mirrors the trends for the DOS at the Fermi level and the e-ph coupling strength under pressure (see Figs.~\ref{F-5} and \ref{F-11}). While our predicted $T_{\rm c}$ values are underestimated compared to the onset superconducting temperatures from the resistivity curves, they show a nice qualitative trend consistent with the experiments.

We also investigated whether the estimated critical temperature is affected when the non-local van der Waals (vdW) functional optB86b~\cite{optB86b, Thonhauser2007, Thonhauser2015, Berland2015, Langreth2009, Sabatini2012} is included in the DFT calculations. In the case of the LP phase, the phonons remain almost unchanged and as a result the $T_{\rm c}$ is unaffected. For the HP phase, the lowest optical phonon branches along the $D$-$B$ direction soften compared to the calculations without vdW. The softening varies from 3~meV at 6~GPa to 1.5~meV at 15~GPa, respectively, and it is due to a small compression along the out-of-plane direction. This led to a modest increase in the e-ph coupling, resulting in a rise in the $T_{\rm c}$ of about 25\% on average, which still falls short of the experimental values. Similar underestimation of the computed $T_{\rm c}$ has been found in other layered compounds under pressure, such as MoTe$_2$~\cite{Paudyal2020} and SnSe~\cite{Marini2019}. The discrepancy between experiments and computations was attributed to 
the sensitivity of the electronic structure to the crystal parameters~\cite{Paudyal2020}, the coexistence of multiple phases~\cite{Marini2019}, and the substantial difference between the measured onset and zero resistance $T_{\rm c}$~\cite{Paudyal2020,Marini2019} under pressure. 

Finally, to gain insight into the anisotropy of the e-ph coupling, we evaluated the momentum-resolved e-ph coupling strength $\lambda_{\textbf{k}}$ for the LP and the HP phase at various pressures. As shown in Fig.~\ref{F-9}(a-b), $\lambda_{\textbf{k}}$ displays a single peak with relatively weak anisotropy in the momentum space. For 6~GPa (HP phase), we also solved the anisotropic full-bandwidth Eliashberg equations~\cite{EPW2023} where the sparse sampling approach with the intermediate representation~\cite{Shinaoka2017,Li2020,Wallerberger2023} was employed to perform the summation over the Matsubara frequencies. We found that the multiple-sheet Fermi surface gives rise to a single anisotropic gap with a distinguished peak at about 0.1~meV in the $T$ = 0~K limit, as shown in Fig.~\ref{F-9}(c). The energy distribution of the superconducting gap reflects closely the anisotropy in $\lambda_{\textbf{k}}$. We obtain an anisotropic $T_{\rm c}$ of 1.1~K for $\mu_c^*=0.10$, a value identical to the one found for the isotropic gap calculation at 6~GPa.

%%%%%%%%%%%%%%%%%%%%%%%%%%

\section{Discussion and conclusions}

Summarizing the obtained experimental data we can conclude that AuAgTe$_4$ undergoes a first order structural phase transition from the LP $P2/c$ to the HP $P2/m$ phase upon pressure at about 5~GPa. This phase transition is manifested by the observed changes in XRD patterns, Raman spectra, dramatic changes of the $R(T)$ behavior and is in excellent agreement with first-principles calculations. We note that there is no sign of hysteresis: upon decompression the LP phase returns at $\sim$5~GPa. The transition to the HP phase is accompanied by a lattice volume contraction of $\sim$4\%. The LP phase is characterized by a significant distortion of the AuTe$_6$ and AgTe$_6$ octahedra due to the linear and second-order Jahn-Teller effect: the six Te atoms form elongated (4 short and 2 long Au/Ag-Te bonds) and compressed (2 short and 4 long Au/Ag-Te bonds) octahedra around the Au and Ag atoms.  These strong Jahn-Teller distortions also give rise to two rather different Te-Te interlayer distances~\cite{Streltsov2018}, see Figs.~\ref{F-1a}(a) and \ref{F-2a}(b).

As usual, with increasing pressure one expects suppression of the Jahn-Teller distortions and indeed the difference between the two sets of Te-Te distances is reduced and then abruptly disappears at $\sim$5~GPa at the transition to the $P2/m$ structure. A similar pressure dependence is also observed for the short and long Au-Te and Ag-Te intralayer bond lengths (Fig.~\ref{F-2a}(b,c)). These changes are due to the sliding of the atomic layers with respect to each other during compression, leading finally to the transition to the $P2/m$ structure, which has regular Te octahedra around Au and Ag atoms (Fig.~\ref{F-1a}(b)).

Our resistivity measurements revealed superconductivity in both AuAgTe$_4$ phases, the LP and HP. However, the LP phase becomes superconducting only  above $\sim$1.5~GPa and shows an almost linear increase of the critical temperature with pressure up to $\sim$6~GPa (with a maximum $T_c \approx 2.5$~K) followed by a slower decrease. The HP $P2/m$ phase, once it occurs, has a higher critical temperature of $\sim$3.5~K ($i.e.$ even higher than in pure AuTe$_2$, with the maximum $T_c \approx 2.3$~K \cite{Kitagawa2013}), and shows a trend to a sluggish $T_c$ decrease under pressure. It is noteworthy that in the case of calaverite AuTe$_2$ the superconductivity had been proposed to be induced by breaking of Te–Te dimers, which exist in the LP $C2/m$ phase, but disappear in the  superconducting HP $P\bar{3}m1$ phase~\cite{Kudo2013}. Alternatively, it was proposed as well that the breaking of the Te–Te dimers is not directly related to the onset of the superconductivity but that the tendency of charge disproportionation of Au$^{2+}$ into, nominally, Au$^{1+}$ and Au$^{3+}$, could be crucial in the formation of Cooper pairs leading to superconductivity under pressure~\cite{Streltsov2018}. Present results show that the situation is even more delicate here, since superconductivity appears already in the low-pressure ``dimerized'' phase. 

Our first-principles computations demonstrate that most probably the superconductivity here is of a conventional type, with the low-energy phonon modes dominating the electron-phonon interactions. Although breaking of the Te–Te dimers is not directly responsible for the onset of superconductivity in AuAgTe$_4$, it results in an appreciable increase in the critical temperature following the transition into the $P2/m$ phase. This is due to an increase in the electronic density of states at the Fermi level related to closing of the pseudogap and to the phonon softening. Overall, our theoretical estimates of the superconducting critical temperature are in good agreement with the experimental results, following a similar trend under applied pressure. These findings can be important not only for silvanite, but also for other similar materials such as puzzling IrTe$_2$. In the case of IrTe$_2$, there is an anomalous structural transition at 270 K, the origin of which is debated, superconductivity induced by intercalation or doping, and complete reconstruction of electronic structure and Ir-Ir dimerization in a monolayer~\cite{Yang2012,li2014,ideta2018,pyon2012,song2021,park2021}.

%\section{Conclusions}
%\textcolor{red}{The conclusions section should come in this section at the end of the article, before the Conflicts of interest statement.}

\section*{Author Contributions}

Yehezkel Amiel: Experiment (Electrical transport measurements), Data analysis, Writing - Review \& Editing; Gyanu P. Kafle: Investigation (DFT Calculations), Formal analysis, Writing - Original Draft, Visualization; Evgenia V. Komleva: Investigation (DFT Calculations), Visualization; Eran Greenberg: Experiment (XRD), Data analysis, Visualization, Writing - Review \& Editing; Yuri S. Ponosov: Experiment (Raman), Writing - Original Draft; Stella Chariton: Experiment (XRD), Data Analysis; Barbara Lavina: Experiment (XRD), Data Analysis; Dongzhou Zhang: Experiment (XRD), Data analysis; Alexander Palevski: Resources, Writing - Review \& Editing; Alexey V. Ushakov: Investigation (DFT Calculations); Hitoshi Mori: Methodology, Software (Code Development); Daniel I. Khomskii: Idea and Conceptualization; Igor I. Mazin, Sergey V. Streltsov, Elena R. Margine: Resources, Conceptualization, Writing - Original Draft \& Review and Editing, Supervision; Gregory Kh. Rozenberg: Conceptualization, Project Supervision, Writing - Original Draft \& Review and Editing.

\section*{Conflicts of interest}
There are no conflicts to declare.

\section*{Acknowledgements}
We are grateful to Prof. L. Bohat\'y and Prof. P. Becker-Bohat\'y for providing us with natural single-crystals. We thank I. Silber and G. Tuvia for assisting with the resistance measurements. 

This work was supported by the National Science Foundation under Grant No. DMR-2035518 (for superconductivity analysis) and Grant No. OAC-2103991 (for code development). This research was supported by the Israel Science Foundation (Grants No. 1552/18 and No. 1748/20). This work used the Expanse system at the San Diego Supercomputer Center via allocation TG-DMR180071 and the Frontera supercomputer at the Texas Advanced Computing Center via the Leadership Resource Allocation (LRAC) award DMR22004. Expanse is supported by the Extreme Science and Engineering Discovery Environment (XSEDE) program~\cite{XSEDE} through NSF Award No. ACI-1548562, and Frontera is supported by NSF Award No. OAC-1818253~\cite{Frontera}.
Portions of this work were performed at GeoSoilEnviroCARS (The University of Chicago, Sector 13), Advanced Photon Source (APS), Argonne National Laboratory. GeoSoilEnviroCARS is supported by the National Science Foundation – Earth Sciences (EAR – 1634415). Use of the COMPRES-GSECARS gas loading system was supported by COMPRES under NSF Cooperative Agreement EAR -1606856 and by GSECARS through NSF grant EAR-1634415 and DOE grant DE-FG02-94ER14466. This research used resources of the Advanced Photon Source, a U.S. Department of Energy (DOE) Office of Science User Facility operated for the DOE Office of Science by Argonne National Laboratory under Contract No. DE-AC02-06CH11357.
E.V.K., Yu.S.P., A.V.U., and S.V.S thank the Russian Ministry of Science and High Education (project ‘‘Quantum’’ No. 122021000038-7).

%%%END OF MAIN TEXT%%%


\balance


%%%REFERENCES%%%
\bibliography{refs} %You need to replace "rsc" on this line with the name of your .bib file
\bibliographystyle{rsc} %the RSC's .bst file

\end{document}